# WebGPU-SPY: Finding Fingerprints in the Sandbox through GPU Cache Attacks

Ethan Ferguson[*]
Binghamton University
Binghamton, NY, USA
efergus3@binghamton.edu

Adam Wilson[*]
Binghamton University
Binghamton, NY, USA
awilso16@binghamton.edu

Hoda Naghibijouybari
Binghamton University
Binghamton, NY, USA
hnaghibi@binghamton.edu

## ABSTRACT
Microarchitectural attacks on CPU structures have been studied in native applications, as well as in web browsers. These attacks continue to be a substantial threat to computing systems at all scales.

With the proliferation of heterogeneous systems and integration of hardware accelerators in every computing system, modern web browsers provide the support of GPU-based acceleration for the graphics and rendering processes. Emerging web standards also support the GPU acceleration of general-purpose computation within web browsers.

In this paper, we present a new attack vector for microarchitectural attacks in web browsers. We use emerging GPU accelerating APIs in modern browsers (specifically WebGPU) to launch a GPU-based cache side channel attack on the compute stack of the GPU that spies on victim activities on the graphics (rendering) stack of the GPU. Unlike prior works that rely on JavaScript APIs or software interfaces to build timing primitives, we build the timer using GPU hardware resources and develop a cache side channel attack on Intel's integrated GPUs. We leverage the GPU's inherent parallelism at different levels to develop high-resolution parallel attacks. We demonstrate that GPU-based cache attacks can achieve a precision of 90% for website fingerprinting of 100 top websites. We also discuss potential countermeasures against the proposed attack to secure the systems at a critical time when these web standards are being developed and before they are widely deployed.

[*]Both authors contributed equally to this research.

## 1 INTRODUCTION
Microarchitectural side channel attacks leak sensitive data through unintended side-effects of program execution observed through shared microarchitectural resources. Attacks have been developed on a variety of microarchitectural structures including different levels of cache [28, 37, 38, 47, 56, 59–61], branch predictors [24], random number generators [23], and others [17, 34]. These attacks continue to be a substantial threat to computing systems at all scales where trusted software can be co-located with untrusted or compromised software on the same hardware resources.

Caches (especially the cross-core shared Last-Level Cache (LLC)) have served as the most important microarchitectural structure to implement timing side channels. In a cache side channel, a spy application first brings the shared cache to a known state, waits for the victim application to execute, and then measures the access time to the shared cache to observe contention from the victim application's activities, which are potentially correlated to its secret data due to some data-dependent software or hardware implementation details. To distinguish the cache hits from cache misses in a cache attack, a high-resolution timer is required.

Although attacks from native applications can be very dangerous, side channels, specifically timing attacks have been recently studied in JavaScript, as well. Most of these attacks use JavaScript's timer [8] to carry out timing measurement [16, 45, 54]. To protect against these timing attacks, all major browsers limited the resolution of the timer [1, 18, 62]. As a result, the timer is not precise enough to distinguish cache hits from misses.

Several works propose some timing primitives in JavaScript to recover high-precision timers [26, 35, 36, 50]. Gras et al. [26] propose two mechanisms (shared memory counter and time to tick) to craft a high-resolution timer in JavaScript. They use a dedicated JavaScript web worker for counting through a shared memory area (SharedArrayBuffers [39] interface) between the main JavaScript thread and the counting web worker. To respond to these attacks, major browser vendors disabled the SharedArrayBuffers interface in JavaScript [4]. This interface has been recently re-enabled for secure contexts only (same-origin) [39]. Some recent works developed coarse-grained cache attacks using the limited low-precision timer in JavaScript. Shusterman et al. [52] use the low-precision timer to implement a cache occupancy channel on LLC; a coarse-grained Prime+Probe attack in which the whole LLC is being probed. Although no spatial information on the victim's accesses can be extracted by cache occupancy channels, these coarse-grained attacks have been shown to be effective for the aggregate measurement of the victim's activities (such as website fingerprinting).

Until recently, most of the microarchitectural side channel attacks have been demonstrated on CPUs. However, modern computing systems are increasingly heterogeneous, combining CPUs along with general-purpose or specialized accelerators to perform application-specific computations offering higher performance and significant power advantages. Graphics Processing Units (GPUs) are the most widely used accelerators with market penetration into a wide range of end-user systems, edge devices, autonomous systems, large-scale HPC clusters, and clouds to enhance the performance of both multimedia and computational workloads. Recent work demonstrated that GPUs in user devices and cloud-based systems are also vulnerable to microarchitectural covert and side channel attacks in native applications [22, 32, 33, 41, 42].

New web standards are increasingly making it possible for web pages to task GPUs on client devices to improve the browsing experience. This includes WebGL [11] which brings GPU-accelerated 3D graphics to the web. Frigo et al. [25] use WebGL timing APIs to implement a Rowhammer attack on DRAM through integrated GPUs in mobile SoCs. In response to this attack, both Chrome and Firefox disabled the WebGL timer [40], which was re-enabled in same-origin later.

Unlike WebGL which is designed only for graphics applications, new GPU-accelerating APIs have emerged to support GPU acceleration of both graphics and computations in modern web browsers. We demonstrate that these emerging GPU-accelerating APIs on the web (such as WebGPU [12] which is called the "future web standard" for both accelerated graphics and compute processes) can offer unique opportunities for attackers. Starting from a malicious webpage, a remote attacker without special access to the system can launch an attack that executes on the compute stack of GPU and spies on the rendering process of a victim user.

In this paper, we study a novel attack vector for microarchitectural side channels in JavaScript through WebGPU which exposes the general-purpose compute capability to attackers, enabling them to spy on the rendering process. We demonstrate that the GPU's unique architecture and capabilities enable an attacker to (1) build a high-resolution timer on hardware resources that can not be easily disabled like software interfaces such as JavaScript timer, SharedArrayBuffer, and WebGL API; (2) develop a low-noise GPU-based cache occupancy channel to spy on the victim's activities (across the compute and graphics stack of the GPU – "cross-stack" attack); and (3) leverage the GPU's inherent parallelism to achieve *high-resolution* cache attacks.

We also discuss potential countermeasures to secure the systems against this class of attacks. Understanding the type of access provided to the attacker through these GPU-accelerated extensions can help us design these emerging interfaces to reduce the threat posed by these attacks, at a critical time when these web standards are being developed and before they are widely deployed.

In summary, the paper makes the following contributions:
- We implement a high-resolution timer using GPU hardware resources in the web browser that does not rely on any JavaScript software interfaces and can bypass all the existing mitigations against microarchitectural attacks in major web browsers.
- We identify a new low-noise leakage vector (the GPU's internal L3 cache) to develop cache attacks in web browsers. The

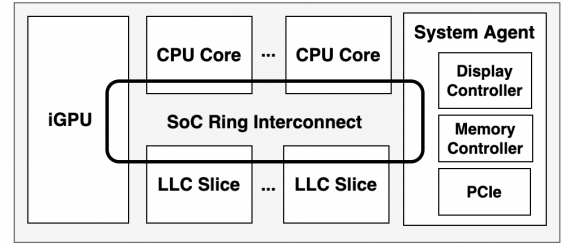

Figure 1: Architecture of Intel SoC

GPU's internal L3 cache serves as the highest (and smallest) level of cache where the compute and rendering accesses interfere, enabling high-resolution cross-stack attacks within the GPU (where the rendering process is executed). It is also not affected by noise from CPU applications.
- We develop a remote cache occupancy channel through the WebGPU API on Intel-based systems and demonstrate an end-to-end website fingerprinting attack. We evaluate the use of two machine learning techniques, Nearest Neighbor and Random Forest, and a deep learning model for fingerprinting websites based on the cache access timing traces collected while being loaded and rendered by the GPU.
- We leverage the GPU's parallelism at different levels to develop parallel attacks and increase the resolution of our cache occupancy channel.

## 2 BACKGROUND

We consider Intel integrated GPUs (iGPUs) in our attacks as the case study, as Intel dominates with more than 60% of the graphics card market [2] in desktop PCs and laptops, given that almost all Intel CPUs include an integrated GPU.

In this section, we introduce the Intel integrated GPU architecture and web-based GPU accelerating APIs to provide the background necessary to understand our attack.

### 2.1 Architecture of Intel Integrated GPUs

Most of Intel's CPUs in laptops and desktops have integrated GPUs on the same die as the CPU, which provides graphics, compute, media, and display capabilities. Integrated GPUs accelerate the real-time rendering process and multimedia heavy workloads without the need for a separate (bulky, expensive, and power-hungry) discrete GPU. Its underlying compute architecture also offers general-purpose compute capabilities with teraFLOPS performance. Figure 1 shows the overall architecture of an Intel System-on-Chip (SoC). The iGPU is on the same die as the CPU and connected to CPU cores and the rest of SoC components with a ring interconnect: a bidirectional 32-byte wide data bus. The CPU and iGPU share the Last-Level Cache (LLC), system agent, and memory subsystem. The system agent bundles the memory management unit, memory controller, display controller, and other I/O controllers [29].

The iGPU consists of a number of slices, each slice has a number of subslices, and each subslice has of a number of Execution Units (EUs). These modular building blocks enable the creation of many product variants with different architectures for a variety of platforms (e.g. Intel HD Graphics, UHD Graphics, Iris Graphics, Iris Plus Graphics, and Iris Xe Graphics). Figure 2 shows the internal architecture of a Gen9 integrated GPU (Intel HD Graphics 530) with a single slice that is composed of three subslices for a total of 24 EUs. Each subslice is equipped with private L1 and L2 sampler caches that are used only in the graphics stack for read-only memory fetch of sampling the textures and image surfaces. There is an L3 cache shared across all subslices and used for both computational and graphics stacks.

The dataport in each subslice is a memory load/store unit that supports efficient read/write operations for a variety of general-purpose buffer accesses and dynamically coalesces scattered memory operations of threads in an SIMD-width of threads into fewer operations over non-duplicated 64-byte cache line requests. All samplers and dataports have their own separate memory interfaces to the L3. In Gen9-based Intel iGPU architectures, the L3 cache size is 768KB per slice, which can be allocated as Shared Local Memory (SLM) or as a data cache. Shared local memory is a highly banked data structure in the L3 complex that supports programmer-managed data for sharing among EU hardware threads within the same subslice. In Gen9-based architectures, the shared local memory size is 64KB per subslice, accessible from all 8 EUs in the subslice. Gen9.5 Graphics architecture introduces some light enhancements.

Unlike Gen9 Graphics architectures, in Gen11 iGPUs [30] the shared local memory (SLM) is integrated close to the EUs. Therefore, SLM and main memory accesses are split and the SLM traffic does not interfere with the L3/memory access through the sampler unit or the dataport. In Section 4, we build our customized timer using SLM in the GPU's hardware and we believe this separated access pathway in Gen11 architecture enables the attacker to build a more reliable and noise-free customized timer in the GPU's SLM without interfering with the cache accesses.

Furthermore, in Gen11-based architectures, 8 subslices (in total 64 EUs) are clustered in a single slice, and the size of the L3 cache is increased to 3 MB for the application data cache and graphics pipeline. Despite the larger L3 cache size in Gen11 architecture, more subslices provide a higher level of parallelism. We develop our attack on three recent iGPU architectures: Gen9, Gen9.5, and Gen11 architectures.

### 2.2 Native GPU Programming APIs

Massively parallel GPUs accelerate graphics workloads, as well as general-purpose computations. In native applications, GPUs are programmed through OpenGL [7] or OpenGL-ES[6] APIs on the graphics stack. The programming language on the computational stack is mostly OpenCL [5] (and CUDA [3] for Nvidia GPUs).

Since our attack is developed using the computational stack of GPUs, in this subsection we provide the necessary background on general-purpose programming on GPUs. GPUs operate in Single-Instruction Multiple Data (SIMD) mode, such that in each cycle, multiple threads are executing the same instruction on multiple data in parallel. Each GPU application consists of some GPU kernels and each GPU kernel launches a large number of threads grouped in workgroups (that are further divided into wavefronts). Workgroups are assigned to subslices in a round-robin manner. Within each subslice, there is a local thread dispatcher, which dispatches the SIMD-width (wavefront) of threads from that workgroup to EUs to be executed in parallel. Shared local memory is shared across all threads within a workgroup and is private to the workgroup.

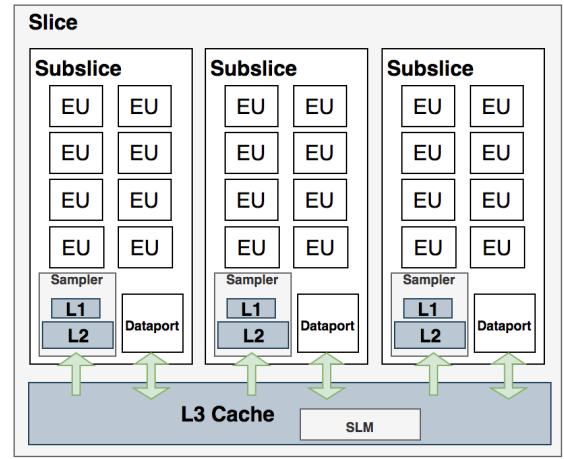

Figure 2: Architecture of Intel graphics Gen9

### 2.3 Web-based GPU Programming APIs

Modern web browsers have started to support GPU-based hardware accelerations for high-performance and efficient rendering processes. This includes WebGL [11] which is based on OpenGL-ES and is a cross-platform API to bring GPU-accelerated 3D graphics to the web.

**WebGPU:** Although WebGL is designed only for graphics applications, the recent trend to integrate both graphics and compute APIs in native applications (e.g. Vulkan [10]) has led to emerging JavaScript APIs such as WebGPU, which is called the "future web standard" for both accelerated graphics and compute. Therefore, WebGPU [12] supports the use of general-purpose compute functionality within web browsers. WebGPU is available today in Chrome 113 on ChromeOS, macOS, and Windows, with other platforms coming soon [13].

A WebGPU-based application is a part of a JavaScript program that is able to launch a compute GPU kernel (called a shader) to be executed on the GPU. Like native general-purpose programs, this GPU shader consists of a number of workgroups (groups of threads) that are assigned to different subslices and will be scheduled on EUs for execution.

### 2.4 GPU-based Rendering in Web Browsers

With GPUs now an integral part of every end-user computing device, web browsers have started to use this underlying hardware more effectively to achieve better performance and power savings in screen rendering. Using the GPU to composite the contents of a web page can result in significant speedups. GPU-based hardware acceleration for rendering is enabled in all major web browsers by default (even without enabling WebGL or WebGPU APIs). The website's content is offloaded to GPU memory, fetched by GPU sampler units (passing through L3 and also L1 and L2 sampler caches), processed on the GPU, and finally dumped in the framebuffer (graphics memory), and rendered on the screen.

This GPU-based rendering process leaves some footprints on GPU's internal microarchitectural structures (e.g. caches) and provides an opportunity for the attackers to spy on users' browsing and other activities through the shared resources within the GPU.

In this paper, we use WebGPU APIs to launch our attack and we show that exposing GPU hardware acceleration for general-purpose computations through APIs such as WebGPU provides a strong leakage vector for developing remote microarchitectural attacks that can spy in the JavaScript sandbox and bypass all existing mitigations against microarchitectural attacks in JavaScript.

## 2.5 Cache-based Side Channel Attacks

Caches have been shown to serve as the most well-known medium for microarchitectural covert and side channel attacks. Many variants of cache-based side channels have been studied in native applications [28, 37, 38, 60, 61], as well as in JavaScript [45, 51, 52].

Depending on the system architecture, system support, and the attacker's ability, several techniques have been studied to develop cache-based side channels. One of the most well-known techniques is the Prime+Probe attack, which is implemented in three steps: (1) the attacker Primes the shared cache by accessing its own data from the memory and filling a specific cache set, (2) the attacker waits for the victim application to do some memory accesses, and (3) the attacker Probes the cache by accessing its own data and measuring the access time. A high latency shows a cache miss, indicating that the victim has accessed this cache set and evicted the attacker's data. However, low latency is interpreted as "no access" from the victim. Usually, the access pattern of the victim process is correlated to some secret information that will be leaked by cache-based side channel attacks.

In Prime+Probe attacks, in addition to having access to a high-resolution timer to track cache hits/misses, the attacker needs to focus on some specific cache set to probe the victim's activity. To achieve this, the pre-attack step will be to reverse-engineer the address mapping and find the eviction set (a group of addresses that are mapped to the same cache set for prime and probe steps). A more coarse-grained Prime+Probe attack is also possible (called a cache occupancy channel), in which the attacker primes and probes the entire cache and monitors the victim's cache activity over the whole cache size. Cache occupancy channels may not be accurate enough for extracting the secret information from some applications (e.g. encryption); however, they have been shown to be very effective in tracking user activities on the web (e.g. website fingerprinting [52]) in which the rendering process leaves a large footprint all over the entire cache. Cache occupancy channels relieve the attackers from finding the eviction sets in restricted environments such as JavaScript with no system support, no notion of pointers, and limited access to high-resolution timers.

In this paper, we first develop a cache occupancy channel within the iGPU through the GPU's internal L3 cache, launched in JavaScript, and then, we show how the GPU's inherent parallelism (and the GPU-based high-resolution timer) enable the attacker to increase the resolution of cache occupancy channels.

## 3 THREAT MODEL

GPU and CPU have their own cache hierarchies. In integrated heterogeneous systems, the CPU and GPU share the LLC and the system memory. Most prior works on remote cache-based side channel attacks in JavaScript target CPU caches (mostly LLC). Modern web browsers provide GPU acceleration for screen rendering. Some recent works develop microarchitectural attacks on the CPU side and use WebGL timing APIs or Javascript timers to target the rendering process on GPU [20, 53, 57]. Unlike prior work, we develop our attack within the GPU where the rendering process is executed (using a WebGPU-based spy also running on GPU).

Our threat model considers end-to-end side channel attacks in JavaScript through WebGPU. We develop a cache occupancy channel attack on the GPU's internal L3 cache. Our attack assumes that the attacker uses WebGPU to launch a process on the *compute* stack of the GPU and spies on the web browser's rendering process that is also accelerated by GPU hardware on the *graphics* stack (it is enabled by default in major web browsers).

A possible scenario where such an attack may be possible is that a remote attacker without special access to the system designs a malicious website using the WebGPU API; once visited by a victim user, the attack gets launched and executed on the GPU and spies on the rendering process of the victim user. Figure 3 shows our threat model in comparison with prior work.

We do not assume any system support such as huge pages (to facilitate extracting eviction sets which are required in Prime+Probe attacks), and also do not rely on any support from JavaScript or WebGPU APIs for the timing mechanisms. We assume all existing and potential JavaScript and browser mitigations against microarchitectural attacks are in place, including disabled JavaScript/WebGL/(potential) WebGPU timers [1, 18, 40, 62], disabled "SharedArrayBuffer" [4], as well as enabled "site-isolation" patch [9] against transient execution attacks in JavaScript. We develop our attack and show its effectiveness in the presence of all these mitigations.

**Experimental Setup:** We developed and validated our attacks on four different Intel-based machines with different GPU generations and models, OSes, and browser versions: (1) a MacBook Pro with i7-6700HQ CPU and HD Graphics 530 (Gen9 Graphics architecture), (2) a MacBook Pro with I5-7360U CPU and Iris Plus Graphics 640 (Gen9 Graphics architecture), (3) a Dell laptop with i7-11800H CPU and UHD Graphics 620 (Gen9.5 Graphics architecture) with Linux OS, and (4) a Dell laptop with I5-11320H CPU and Iris Xe Graphics (Gen11 Graphics architecture) with Windows 11 OS. We use WebGPU in the experimental versions of Google Chrome, specifically, Google Chrome Canary on macOS and Windows and Google Chrome Dev on the Linux-based system to develop our attack. To cover all Graphics architectures with different OSes, we report the classification results for one Gen9 Graphics architecture (first configuration on macOS), one Gen9.5 Graphics architecture (third configuration on Linux), and one Gen11 Graphics architecture (fourth configuration on Windows).

Current integrated GPUs do not support running multiple computation kernels from separate contexts concurrently and therefore no noise is expected on the compute stack of the GPU. Unlike prior work, our attack is also not affected by the noise on the CPU side,

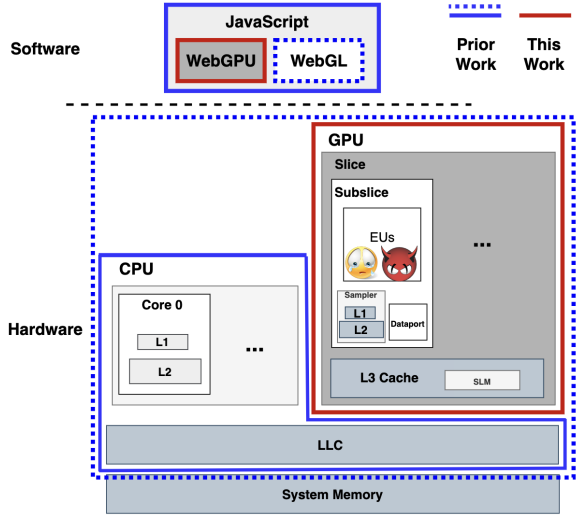

Figure 3: Our threat model (compared with prior work)

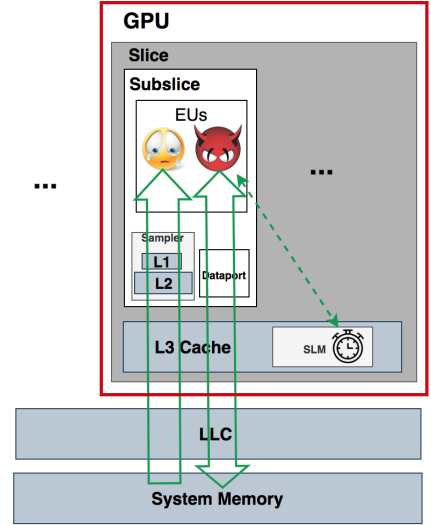

Figure 4: The separated pathways of the Spy (compute) and Victim (graphics) accesses (interference at the L3 cache)

as it is completely developed within the GPU.

**Comparison with prior work:** Until recent years, all JavaScript-based microarchitectural attacks have been developed on the CPU side (targeting large LLC or DRAM) and have used JavaScript software interfaces to build high-resolution timers [26, 45, 50–52].

Most GPU-based microarchitectural attacks are developed in native code [22, 32, 33, 42]. Several recent works proposed GPU-based attacks in browsers that create contention on the graphics stack of GPU and measure the time of GPU tasks on the CPU side using software interfaces (WebGL timer or JavaScript timer) [20, 53, 57].

As shown in Figure 3, our attack is developed completely within the GPU where the rendering is processed. We uniquely identify the GPU's internal L3 cache as the highest and smallest level of cache where the compute and rendering accesses interfere (details in Section 4). Our cross-stack attack uses the emerging general-purpose *compute* capability for the web (supported only by WebGPU in JavaScript) to build the timer using GPU hardware resources (details in Section 4) and spy on the *rendering* process within the GPU. Our attack does not rely on any software interfaces/APIs for building the timer. It is not impacted by the noise from other processes on the CPU side. Moreover, any noise generated by the memory accesses of the attack code itself is isolated from the rendering process (victim) since the spy is on the compute stack of the GPU (details in Section 4). By leveraging the GPU's parallelism and our customized high-resolution timer, we increase the resolution of our cache occupancy channel (details in Section 7).

Table 1 compares our threat model, our attack leakage resource, and our high-resolution timing mechanism with the most relevant microarchitectural attacks in web browsers or GPU-based attacks in native code targeting web browsers.

## 4 ATTACK PRIMITIVES

In this section, we present our attack vector while explaining the primary ingredients necessary for our attack.

### 4.1 Why GPU's Internal L3 Cache?

As discussed in earlier sections, Intel iGPUs have three levels of internal cache, as well as LLC that is shared with the CPU. As shown in Figure 4, L1 and L2 are read-only sampler caches, private to each subslice and dedicated to the graphics stack. The GPU's internal L3 cache is shared across all subslices within the slice and supports caching all memory accesses from the graphics stack (L1 and L2), as well as read/write memory accesses from the compute stack (routed through the dataport). We build our attack on the GPU's L3 cache for the following reasons:

- GPU's L3 cache is the highest level of cache and the smallest one on which the graphics and compute memory accesses interfere. We exploited these separate memory access pathways to monitor the screen rendering process through the L3 cache using a compute spy.
- Since memory requests from the compute spy are directly served by L3, to develop side channel attacks on L3 from a compute context, the attacker does not need to evict any higher level of caches, and it simplifies the attack significantly.
- Unlike the LLC, the GPU's internal L3 cache is not affected by the noise from the processes running on the CPU cores.
- Prior work [22] has shown that the GPU's L3 cache address mapping does not include index hashing, which simplifies our high-resolution parallel attacks (presented in Section 7), in which each thread probes a portion of cache (all in parallel).

**L3 Cache Structure:** In Gen9-based Intel iGPU architectures, The L3 size is 768 KB per slice that can be allocated as application L3

Table 1: Microarchitectural attacks in web browsers or (from GPU native code targeting web browsers)

| | Attack | Processor | Leakage resource | Timing mechanism |
|---|---|---|---|---|
| **Oren et al. [45]** | Side channel | CPU | LLC (Prime+Probe) | JavaScript Timer (Software) |
| **Gras et al. [26]** | Side channel | CPU | LLC (Prime+Probe & Evict+Time) | SharedArrayBuffer (Software) |
| **Schwarz et al. [50]** | Covert channel | CPU | DRAM | SharedArrayBuffer (Software) |
| **Shusterman et al. [51]** | Side channel | CPU | LLC (Cache Occupancy) | JavaScript Timer (Software) |
| **Shusterman et al. [52]** | Side channel | CPU | LLC (Cache Occupancy) | JavaScript Timer (Software) |
| **Wu et al. [57]** | Side channel | from CPU | All rendering resources (CPU,GPU,and screen buffer) | JavaScript timer (Software) |
| **Naghibijouybari et al. [42]** | Side channel | in GPU (from Graphics/Compute **native** code to browsers) | GPU's memory allocation & Performance counter | – (* no timing *) |
| **Frigo et al. [25]** | Rowhammer | from iGPU (Graphics) to DRAM | DRAM | WebGL timer (Software) |
| **Cronin et al.. [20]** | Side channel | from CPU to GPU | GPU contention using WebGL | JavaScript timer (Software) |
| **Laperdix et al. [53]** | Side channel | from CPU to GPU | GPU's EUs | WebGL timer (Software) |
| **WebGPU-SPY (This work)** | Side channel | in iGPU (cross-stack) (from Compute to Graphics stack) | GPU's L3 cache (Cache Occupancy) | Hardware-based timer within GPU |

data cache, buffer for graphics pipeline, or shared local memory (SLM). The typical allocation for application data cache is 512 KB per slice. The L3 cache size is increased to 3MB in Gen11 iGPU architectures.

Shared local memory is integrated into the L3 fabric with the size of 64KB per subslice. Based on Intel documentation [29, 30] and reverse engineering results of prior work [22], The L3 is 64-way set-associative, partitioned into 4 banks (2 bits for address mapping) of each 128 KB. Each bank is further partitioned into 8 sub-banks (3 bits for address mapping) and each sub-bank has 32 sets which require 5 bits in the address bits for address mapping. As a result, a total of 10 bits (5 bits for cache set + 2 bits for cache bank + 3 bits for sub-banks) determine the set of the L3 in which a cache line resides. These bits are the least significant bits of the address after accounting for the cache line offset bits–that is, there is no index hashing in L3. This simplifies the cache attacks in a restricted JavaScript environment with no notion of pointers and no information about the cache address mappings. An attacker with a single thread can probe the whole buffer of L3 cache size to implement a cache occupancy channel, and in an optimized scenario, an attacker can leverage GPU parallelism and launch multiple threads, each probing a portion of the whole buffer, all in parallel to increase the resolution of the attack.

## 4.2 Building High-Resolution Timer

Our attacks do not rely on JavaScript, WebGL, or (potential) WebGPU timing interfaces. We also do not assume the support of software interfaces like "SharedArrayBuffer" or "OffScreenCanvas" in JavaScript that facilitate the building of high-precision timers in the absence of timing APIs, as all of these interfaces have been already disabled or restricted in response to prior microarchitectural attacks in JavaScript.

Instead, we build our customized high-resolution timer by exploiting the GPU's shared local memory in hardware that cannot easily be disabled like software interfaces (e.g. SharedArrayBuffer) as a quick patch. All GPUs have shared memory in hardware, which is a critical resource to enable massively parallel computations, and disabling access to this shared memory leads to a large performance overhead.

We leverage the idea proposed in [22] that built a customized timer for a native CPU-GPU attack. We show that WebGPU exposes this high-resolution hardware-based timer by enabling general-purpose computation in JavaScript. To build the timer, we define a counter value stored in the shared local memory that is private to a subslice and can be accessed by all threads of a workgroup that is assigned to that subslice. We launch a workgroup of 96 threads, in which 64 threads (64/32 = 2 wavefronts, we call these *"counting threads"*) increment the counter value in the shared memory and just 1 single thread in the third wavefront (we call it the *"attacker's thread"* is active and accesses the memory for priming/probing the cache. All these three wavefronts execute in parallel. To measure the access time, the attacker's thread reads the counter value as timestamps before and after the access. Figure 5 shows the implementation of our customized timer on the GPU's shared local memory.

Note that the maximum number of threads per workgroup is 256. Unlike [22] which uses one wavefront as the attacker's wavefront and 224 remaining threads (7 wavefronts) as counting threads, we have observed that 64 threads for counting achieve a reasonable timing accuracy to track cache hits/misses. In Section 7, we discuss how saving the remaining threads helps us speed up the attack.

Despite the limited programming support on the current experimental versions of WebGPU (e.g. lack of volatile memory type and proper barrier instructions to synchronize the threads within a workgroup), we were able to build the customized timer on the compute stack of GPU using WebGPU that is precise enough to determine the L3 cache hits vs. misses, as shown in the histogram in Figure 6. We also observed that the separate access pathways to SLM and the L3 cache in Gen11 architectures provide a more reliable timer with a lower noise level.

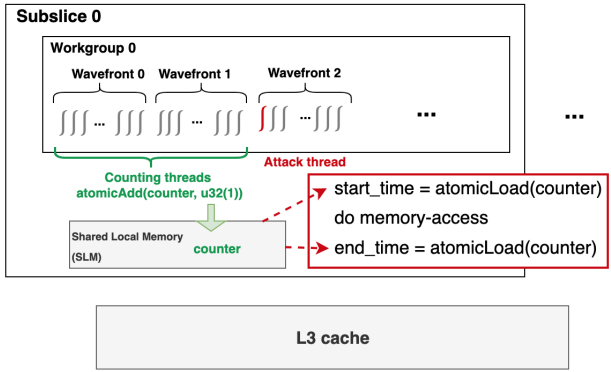

Figure 5: Customized timer on WebGPU

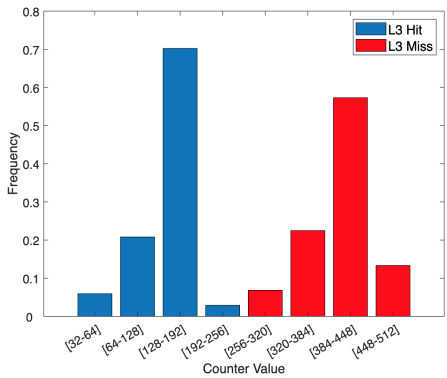

Figure 6: Customized timer characterization in Gen11 iGPU

## 5 CACHE OCCUPANCY CHANNEL

In this section, we develop a WebGPU-based cache occupancy side channel attack. Cache occupancy has previously been implemented on the LLC in native code for covert channels and measuring co-resident activities [19, 49], as well as in JavaScript on the CPU side [52], which has been shown to be very effective for conducting website fingerprinting.

We demonstrate a novel attack vector through the iGPU which is (1) faster, given that the GPU's L3 cache is smaller than LLC (e.g. in our target Gen9-based system, the L3 is 1/16 of the LLC size), (2) has higher quality, since the attack is developed within the GPU where the rendering is processed and is not impacted by the noise from the CPU applications, and (3) achieve higher resolution, since the attacker can build parallel attacks by leveraging GPU parallelism at different levels and high-resolution timer that is built by GPU's hardware resources.

In our cache occupancy channel, the attacker allocates a buffer the size of the GPU's L3 cache (e.g. 256KB in our Gen9 system) and launches a GPU compute shader with 1 single workgroup. This workgroup consists of 3 wavefronts (96 threads), in which 64 threads (first two wavefronts) are used for building the timer and just 1 single thread (the attacker thread) in the third wavefront is active, as shown in Figure 5. The attacker thread accesses the whole buffer repeatedly in a loop and measures the access time. The victim's memory accesses evict the attacker's buffer from the cache, introducing delays for the attacker's next access. Thus, the time to access the attacker's buffer is roughly proportional to the number of cache lines that the victim uses.

Due to instruction-level parallelism (ILP) on the GPU, we observed that the timing instructions may overlap with their previous instructions and even return before the previous instruction finishes. To lower the effect of instruction-level parallelism on GPU and ensure the in-order execution of the critical part of the code, we used some dummy instructions that are dependent on the critical memory access and timing instructions, making the timing measurement instructions artificially dependent on the memory access instructions. We also ensure that these dummy instructions do not interfere with either timing or memory accesses, and as a result, do not add noise to our experiments. We access the buffer in a pointer-chasing manner and randomly permute the order of elements in the buffer to minimize the cache prefetching impact.

To validate our attack, we launch the attack to collect the cache access timing traces when a user is visiting some websites. We call the cache access timing trace a "memorygram" (like prior work[45, 52]).

**Our Observations:** We collected the timing measurements over 5 seconds for each visit to each website. However, we noticed that the GPU shader program of our cache occupancy channel that is repeatedly accessing the large array saturates the GPU resources, and is long enough for the iGPU that the victim user can notice the rendering slowdown. To overcome this challenge, we launch hundreds of very short GPU shader programs, each accessing the whole buffer once and recording the time (that was measured inside the GPU shader using our customized timer and does not include the shader launching overhead). Although it has the overhead of launching a shader program for each point of memorygram and lowers the resolution (sampling rate) of the attack, it does not impact the rendering process of the victim. In Section 7, we show how we can leverage GPU parallelism to increase the resolution of our cache occupancy channel.

Figure 7 shows the extracted memorygrams by our cache occupancy channel using 1 single attacker thread on a single subslice. The X-axis is the iteration number of probing the buffer over time (256 iterations), and the Y-axis is the access time (counter value) of the whole buffer. The Y-axis for all plots ranges from 80000 to 120000. We observe that (1) every website has a unique memorygram due to the different number of objects and different sizes of objects being rendered, and (2) memorygrams of each website are similar across several trials.

## 6 WEBSITE FINGERPRINTING

To evaluate the effectiveness of our GPU-based cache occupancy channel, we implement a website fingerprinting attack on the Alexa top 100 websites. In such an attack scenario, a victim visits different websites in the web browser and the attacker's goal is to find out which websites the victim has visited (through the cache occupancy channel). In this section, we describe our data collection and machine-learning based classification approaches.

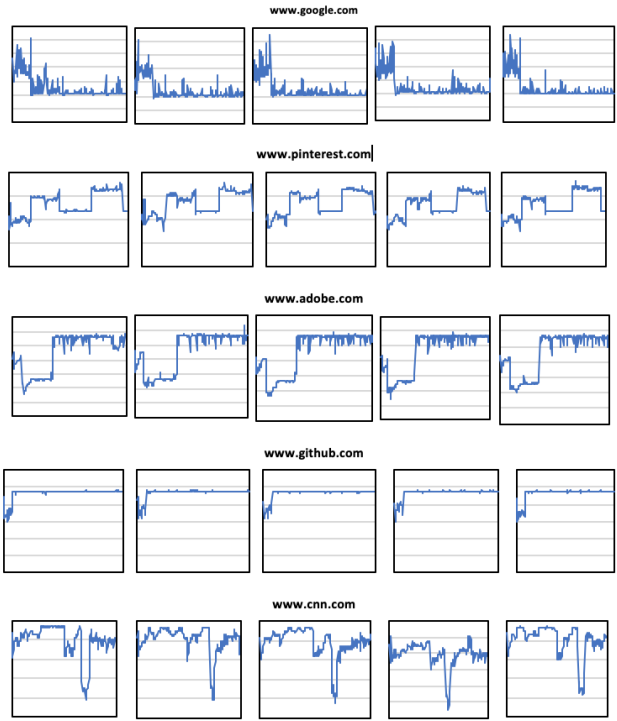

Figure 7: Memorygram of 5 websites (5 visits of each)

## 6.1 Data Collection

We evaluate the website fingerprinting attack on the front pages of the top 100 websites ranked by Alexa. Our GPU-based cache occupancy channel runs on a web browser tab in Google Chrome Canary or Google Chrome Dev with the "WebGPU" flag enabled. We open a website in another tab and we collect the memorygram traces for 5 seconds. The website is left open for an additional 10 seconds, after which the tab is closed and a new tab opens and displays a new website. We visit the top 100 websites 100 times each and collect the memorygrams (in total 10000 memorygrams on each machine). A memorygram is the trace of the cache access latency measured over a given time period, while the website is being rendered. The sampling rate for this basic attack is 50Hz and the average attack time is 5 seconds.

## 6.2 Machine learning based classification

After collecting the memorygram samples, we need to train a Machine Learning (ML) classifier to identify the different websites based on the memorygram time series. The base accuracy rate of this prediction method is 1% for website fingerprinting of 100 websites (closed-world setting). We construct features from the full time series signal and use traditional machine learning classification, similar to [42]. We also evaluate our attack using a Deep Learning (DL) based classifier, similar to some of prior works [51, 52, 57].

In particular, for the ML-classifiers, we computed seven statistical features for the memorygrams collected through the cache occupancy channel, including minimum, maximum, mean, standard deviation, skew, and kurtosis. These features are easy to compute and capture the essence of the distribution of the time series values. The skew and kurtosis capture the shape of the distribution of the time series. Skew characterizes the degree of asymmetry of values, while kurtosis measures the relative flatness of the distribution relative to a normal distribution. We computed these features separately for the first and the second half of the time series recorded for each website. We further divided the data in each half into 4 equal segments and measured the 6 features for each segment as well. This process resulted in the feature set consisting of 60 features. We then used these features to build the classification models based on two standard machine learning algorithms, namely K Nearest Neighbor (KNN) and Random Forest (RF). We also validated the classification models using the standard 10-fold cross-validation method.

For the DL-based classifier, we used the open-source LSTM model and methodology from prior work [51]. In this case, the feature extraction was done inside the neural network and did not require additional preprocessing steps.

Table 2: Classification results of cache occupancy channel

|  | Classifier | Precision | F1-Score | Recall |
|---|---|---|---|---|
| **Gen9** | KNN | 69.3% | 68.2% | 68.2% |
|  | RF | 72.3% | 71.9% | 71.1% |
|  | LSTM | 70.8% | 71% | 71.2% |
| **Gen9.5** | KNN | 68.4% | 69.2% | 69% |
|  | RF | 71.6% | 71% | 71.3% |
|  | LSTM | 69.2% | 70.3% | 70.3% |
| **Gen11** | KNN | 70.4% | 70.2% | 70.9% |
|  | RF | 73.1% | 72.8% | 71.8% |
|  | LSTM | 71.1% | 71% | 70% |

To evaluate the performance of our classifiers, we computed the precision (*Prec*), recall (*Rec*), and F1-Score (*F1*). *Prec* and *Rec* refer to the accuracy of the model in rejecting the negative classes and in accepting positive classes, respectively. Low recall leads to high rejection of positive instances (false negatives) while low precision leads to high acceptance of negative instances (false positives). *F1* represents a balance between precision and recall. Table 2 shows the classification results on 3 machines with 3 different generations of iGPU, CPU, and OSes (the details of configurations are in Section 3) in website fingerprinting of the Alexa top 100 websites.

## 7 PARALLEL ATTACKS

In the basic version of the attack (Section 5), in addition to 64 timing threads, the attacker launches one single extra thread to access the whole buffer of L3 size and measures the access time in a specific time period (5 seconds). The attacker process uses only one subslice of the GPU to spy on the victim process and needs to launch one GPU shader for each probe, which lowers the resolution of our attack. In this subsection, we optimize our cache occupancy channel by leveraging massive GPU parallelism.

As shown in Figure 8, we introduce 3-level parallelism to speed up the cache occupancy channel and monitor the victim's activity with a higher resolution:

**A) Thread_level parallelism:** Each wavefront has 32 parallel threads. We can take advantage of thread-level parallelism on the GPU to have all 32 threads of each wavefront active during the probe process. Therefore, every single thread in the attacker's wavefront is responsible for probing 1/32 of the whole buffer.

**B) Wavefront_level parallelism:** As discussed in earlier sections, shared local memory is private to a workgroup within the subslice, and the maximum number of threads in a workgroup is 256 threads. Since we used just two wavefronts (64 threads) for counting, the remaining 6 wavefronts can be active and utilized as the attacker's wavefronts (each with 32 threads), further speeding up the attack.

**C) Workgroup_level Parallelism:** In another level of parallelism, the iGPU has more than 1 subslice (specifically 3 subslices in Gen9 and Gen9.5 architectures and 8 subslices in Gen11 architectures). We can take advantage of this to further speed up our attack. For example, we launch 3 workgroups (that are assigned to three subslices in a round-robin fashion) and build one instance of the timer (using 64 threads of each workgroup) in each subslice. The remaining threads/wavefronts of each workgroup are responsible for probing a portion of the L3 cache size buffer (e.g. 1/3 of buffer size in Gen9).

By leveraging all three levels of the aforementioned optimization approaches, *ideally* 1 single attacker thread is responsible for probing 1/(# of subslices * 6 wavefronts * 32 threads within a wavefront) of the whole buffer of L3 size in each launch of a GPU shader program: that will be 1/(3*6*32) of 256KB in Gen9 and Gen9.5 architectures. This gives us higher sample rates and as a result, a higher resolution cache occupancy channel on GPU's L3 cache.

**Our Observations:** We first experimented with the thread_level parallelism. We launched the GPU shader with a single workgroup with 3 wavefronts for a total of 96 threads. We increased the number of active threads in the attacker's wavefront one by one and studied its impact on the memorygram traces. We observed an increase in the resolution of memorygrams (monitoring the access latency more frequently in the same period of time) without adding noise for up to 8 active threads (as shown in Figure 9). However, we observed that leveraging more than 8 active threads added noise to the collected memorygrams. This effect is due to the limited number of Execution Units (EUs) in each subslice (8 EUs per subslice). Within an EU, memory operations are all dispatched via "send instructions" that are executed by the "send unit", and each EU is equipped with just one single send unit. If the attacker has more than 8 threads, the self-conflict and contention between these threads to be scheduled on the EUs and send units will add delay to the execution and lead to a higher noise level.

This effect also can be seen in wavefront_level parallelism within the subslice. Ideally, we can launch 8 wavefronts (256 threads) in a workgroup, use 2 among them for timing, and have 6 remaining wavefronts, each with 32 threads for probing the L3 cache. However, the limited number of EUs within each subslice limits the attacker's ability. In the best-case scenario, to achieve a low-noise and high-resolution attack, the attacker can have up to 8 active threads in each subslice. It can be organized in several configurations (e.g. 4 wavefronts each with 2 active threads, or 1 wavefront with 8 active threads, ...).

Since each subslice has its own shared local memory and 8 EUs, we can take full advantage of workgroup_level parallelism. To achieve this, in Gen9-based architectures with 3 subslices, the attacker needs to launch 3 workgroups, each with one instance of the timer within the subslice and 8 active attack threads. Each workgroup probes 1/3 of the L3 size buffer, as a result, each thread within that workgroup probes 1/(3*8) of the buffer (all in parallel). These parallel attacks enable the attacker to probe the whole L3 cache at a higher rate and lead to a higher resolution of memorygrams in a given time period. Our parallel attack could achieve up to a 170Hz sampling rate in an average attack time of 3 seconds.

## 7.1 Website Fingerprinting using Parallel Attacks

Based on our observations, we evaluate two versions of our parallel cache occupancy channels for website fingerprinting: thread_level parallel attack (1 workgroup with 8 active threads) and also workgroup_level + thread_level parallel attack (3 workgroups in Gen9 and Gen9.5 and 8 workgroups in Gen11, each with 8 active threads). In this subsection, we report the traditional ML and LSTM classification results for both of these two parallel attacks on three machines. Similar to the ML classification for the simple attack in Section 6, we construct seven statistical features for the memorygram collected through the parallel channels on Alexa 100 top websites. These features are minimum, maximum, mean, standard deviation, skew, and kurtosis.

**Table 3: Classification results for thread_level parallel cache occupancy channel**

|         | Classifier | Precision | F1-Score | Recall |
|---------|------------|-----------|----------|--------|
| **Gen9**    | KNN  | 80.4% | 79.6% | 79.2% |
|         | RF   | 83.5% | 82.5% | 81.4% |
|         | LSTM | 84.9% | 84.5% | 84.8% |
| **Gen9.5**  | KNN  | 80.1% | 79%   | 79.4% |
|         | RF   | 83.6% | 83.1% | 83%   |
|         | LSTM | 83.4% | 83%   | 83%   |
| **Gen11**   | KNN  | 79.8% | 79.6% | 78.8% |
|         | RF   | 81.3% | 81%   | 81.2% |
|         | LSTM | 82%   | 81.8% | 82%   |

For the ML classifiers, we computed the features separately for the first and the second half of the time series recorded for each website. We further divided the data in each half into 8 equal segments, and measured the 6 features for each segment as well, resulting in 108 features.

Table 3 and Table 4 show the precision (*Prec*), recall (*Rec*), and F1-Score (*F1*) of the classification models for thread_level parallel attack and workgroup_level+thread_level parallel attack, respectively (all for the Alexa top 100 websites). Although the traditional ML classifiers outperformed the LSTM model (Section 6) for our basic attack, we observed better classification results using the LSTM classifier in our high-resolution parallel attacks.

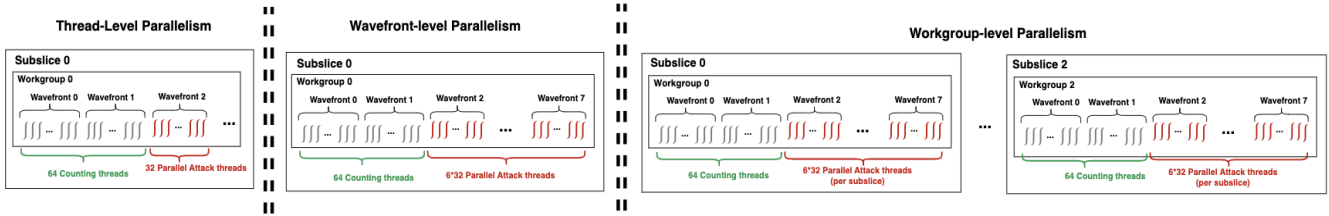

Figure 8: Three levels of GPU parallelism

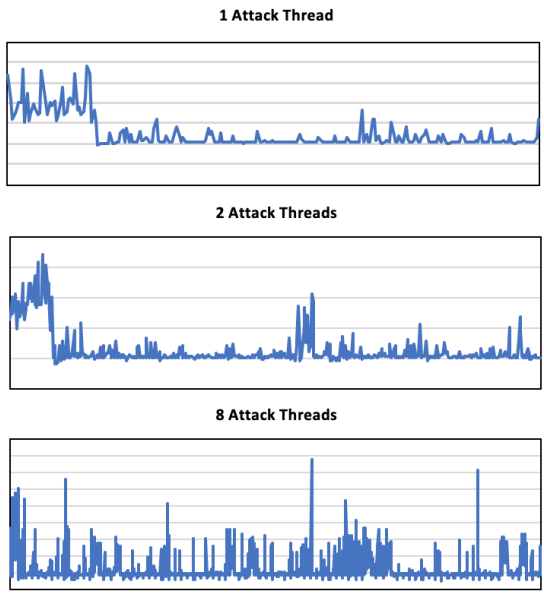

Figure 9: Memorygrams of "www.google.com" website with thread_level parallel attack.

Table 4: Classification results for workgroup_level + thread_level parallel cache occupancy channel

|  | Classifier | Precision | F1-Score | Recall |
|---|---|---|---|---|
| **Gen9** | KNN | 82.3% | 80.1% | 79.8% |
|  | RF | 85.2% | 84.9% | 84.5% |
|  | LSTM | 86.6% | 86.2% | 86% |
| **Gen9.5** | KNN | 81.8% | 81% | 81% |
|  | RF | 86.4% | 86% | 86% |
|  | LSTM | 87.4% | 87.2% | 87.2% |
| **Gen11** | KNN | 86.3% | 86.1% | 86% |
|  | RF | 88.4% | 88.1% | 88.1% |
|  | LSTM | 90.6% | 90.4% | 90.4% |

### 7.2 Sensitivity to browser window size

We considered full-screen browser windows in our experiments in Sections 6 and 7. We also checked if the attack generalizes across window sizes, given that different users may have different size browser windows. We observed that changing the window size results in a similar signal for most websites, and for responsive websites that have dynamic content or do not scale with window size, there is some variance in memorygrams. We trained the LSTM model using full-screen data collected by workgroup_level+thread_level parallel attack on Gen11 architecture and tested with window sizes of 50% and 70% of the screen width. We observed average accuracy of 78% and 84%, respectively using the LSTM classifier. We believe performance can also be improved by training with measurements taken at different window sizes.

### 7.3 Evaluation in Open-World Setting

The reported classification results in previous sections were all on the datasets collected in a closed-world setting (100 traces of every Alexa top 100 website). We also wanted to check the accuracy of our attack in an open-world setting. For this, we collected the traces following the methodology stated in prior works [52, 57]. In this setting, the attacker monitors access to a set of sensitive websites, and is expected to classify them with high accuracy. Additionally, there is a large set of non-sensitive web pages, all of which the attacker is expected to generally label as "non-sensitive" [52]. Datasets in this setting consist of the closed-world dataset (containing 10000 traces, 100 traces for each website) plus 5000 other websites. The base accuracy rate of this prediction method is about 33%. Since LSTM classification performed better on our parallel attacks, we evaluated our workgroup_level+thread_level parallel attack using LSTM in open-world setting. The classification results are shown in Table 5.

Table 5: Classification results of workgroup_level + thread_level parallel attack in Open-World Setting

|  | Classifier | Precision | F1-Score | Recall |
|---|---|---|---|---|
| **Gen9** | LSTM | 87.3% | 87.3% | 87.4% |
| **Gen9.5** | LSTM | 86.5% | 86.1% | 86.1% |
| **Gen11** | LSTM | 89.8% | 89.8% | 89.6% |

## 8 POSSIBLE DEFENSES

Our cache occupancy channel attack in this paper introduces a new attack vector for microarchitectural attacks in JavaScript that are enabled by emerging GPU-accelerating web interfaces. We believe the secure design of these web standards is very important at this time before they are widely deployed in critical computing platforms and application domains.

To defend against this new class of attacks, mitigations can be designed at different levels of software and hardware: (1) at the microarchitectural level, (2) at JavaScript runtime, accelerated extensions, and permission models, and (3) at the GPU driver level.

At the hardware and microarchitectural level, we believe classes of defenses that have been developed against other microarchitectural covert and side channels could potentially be applied to mitigate our attack. The most well-known technique is static or dynamic partitioning of caches[21, 31, 46, 48, 58]. Since we build our customized timer using hardware resources (shared local memory) available on GPU, although disabling the timer or restricting access to it is not straightforward, reducing the resolution of the timer, or adding noise to the timer can be possible by limiting the throughput of atomic operation execution.

Some potential mitigations at the JavaScript and GPU driver levels include:

**Heuristic profiling in JavaScript to detect and prevent the attack:** Our attack accesses memory in a very particular pattern that can be detected. As part of optimization mechanisms, modern JavaScript runtime already analyzes the runtime performance of code, thus it could be possible for the JavaScript runtime to detect profiling-like behavior from executing code, and modify its response accordingly (e.g., by temporal or spatial partitioning or applying some permission restrictions in accelerated extensions).

Note that any defense design on JavaScript has to be reflected or applied to the WebGPU API as well, to effectively mitigate our proposed attack on GPU.

**Partitioning through the GPU device driver:** The approach partitions the resources (such as L3 cache sets/ways) between security domains for different processes. With OS support, the device driver will be provided with information on memory allocations and memory mapping. Thus during the GPU process initialization in JavaScript, before allocating any GPU memory for an application, the driver assigns some dedicated partitions of GPU caches to a given process, and all future memory allocations for that process will be mapped to that partition.

This approach will be uniquely possible in CPU-GPU systems. While a CPU program may access millions of memory objects, the unique programming model and disciplined memory model on GPUs limit the number of buffers used in a GPU kernel. The programmers also have to provide detailed information about memory buffers, such as size and read-only attributes. As opposed to CPU memory objects which can be freely allocated and deallocated during runtime, GPU memory buffers are mostly allocated before a shader program launches and deallocated after completion. This makes the partitioning feasible at the driver level in the GPU initialization step (before the GPU shader program execution).

## 9 RELATED WORK

In this section, we organize the discussion of the most related work into two categories: (1) microarchitectural attacks and defenses in JavaScript, and (2) GPU-based microarchitectural attacks in native code. Then we compare the classification accuracy of our attack with prior website fingerprinting attacks in both categories.

### 9.1 Microarchitectural Attacks and Defenses in JavaScript

Many prior works studied timing side channel attacks in JavaScript. In this subsection, we review only the *microarchitectural* side channel attacks in JavaScript and those works that built high-resolution timers in JavaScript, since these are the most related works to our attack.

Oren et al. [45] implemented a JavaScript-based Prime+Probe attack on the last level cache on Intel CPUs to spy on the user's mouse movements and network activities. To protect against this timing attack, all major browsers limited the resolution of the timer [1, 18, 62]. This low-precision timer has been shown to be sufficient for conducting some types of attacks. Gruss et al. [27] proposed a memory page deduplication timing attack to identify which websites the user currently has open. However, to distinguish cache hits from cache misses in a cache attack, a high-resolution timer is required. Several works proposed some timing primitives in JavaScript to recover highly accurate timestamps [26, 35, 36, 50]. Lipp et al. [36] proposed a keystroke interrupt-timing attack implemented in JavaScript using a counter as a high-resolution timer. Kohlbrenner et al. [35] studied the clock-edge technique. Gras et al. [26] proposed two mechanisms (shared memory counter and time to tick) to craft a high-resolution timer in JavaScript. They used a dedicated JavaScript web worker for counting through a shared memory area (SharedArrayBuffers [39] interface) between the main JavaScript thread and the counting web worker. Schwarz et al. [50] also used similar techniques to build a high-resolution timer and implement a new DRAM-based covert channel between a website and an unprivileged app. To respond to these attacks, major browser vendors disabled the SharedArrayBuffers interface in JavaScript [4] which has been recently re-enabled for same-origin only.

Some recent works developed coarse-grained cache attacks using the limited low-precision timer in JavaScript. Shusterman et al. [52] used the low-precision timer to implement a cache occupancy channel on LLC. The same group [51] also developed cache occupancy channels without using JavaScript features, instead using DNS response time as a timer. All of these attack models are on the CPU, providing a different attack vector than our threat model.

In recent years, some timing side channels have been proposed on the graphics stack of the GPU in JavaScript. Frigo et al. [25] used WebGL timing APIs to implement a Rowhammer attack on integrated GPUs in mobile SoCs. They used the WebGL timer to find the contiguous areas of physical memory to conduct the Rowhammer attack. As discussed earlier, in response to this attack, both Chrome and Firefox disabled the WebGL timer [40], which is re-enabled for secure contexts only. Laor et al. [53] presented a device identification technique that uses the WebGL API in JavaScript to access the GPU and execute a sequence of drawing operations. The attacker measures the speed of each EU as a fingerprint. They use the (re-enabled) WebGL timer to measure the time of drawing commands. Cronin et al. [20] developed website fingerprinting using GPU contention attack from CPU in ARM SoCs. They used WebGL to launch GPU kernels and measured the time on the CPU side through JavaScript timer. Wu et al. [57] also used the low-precision JavaScript timer to measure the aggregate contention on all rendering resources (on the CPU, GPU, and screen buffer). This attack is also developed on the CPU and targets the rendering process. Our WebGPU-SPY attack has the same average attack time of 3 seconds as [57], with a higher sampling rate (up to 170Hz in our parallel attack compared with 10-60Hz in [57]).

Table 6: Classification accuracy comparison of website-fingerprinting attacks on Alexa top websites

|  | Attack | Accuracy (%) | # of Websites |
|---|---|---|---|
| **Naghibijouybari et al. [42]** | *native code* side channel (GPU memory API and perf. counters) | 90.4 - 93 (94) | 200 (100) |
| **Oren et al. [45]** | side channel (LLC) | 88.6 | 8 |
| **Shusterman et al. [52]** | side channel (LLC) | 70-90 | 100 |
| **Shusterman et al. [51]** | side channel (LLC) | 87.5 | 100 |
| **Wu et al. [57]** | side channel (from CPU to all rendering resources) | up to 88 | 100 |
| **Cronin et al. [20]** | side channel (from CPU to GPU) | up to 90.3 | 100 |
| **WebGPU-SPY (This work)** | side channel (GPU's L3 cache) | up to 90.6 | 100 |

In Section 3 (Table 1), we compared our threat model with prior microarchitectural attacks in JavaScript.

## 9.2 Microarchitectural Attacks on GPUs (native code)

Microarchitectural side-channel attacks have been extensively studied in native code on different resources on CPUs. Some recent works demonstrate that GPUs are also vulnerable to microarchitectural covert and side-channel attacks. Most of these works have been proposed on discrete GPUs with dedicated memory. Jiang et al. [32, 33] present architectural timing side channel attacks on GPUs by exploiting key-dependent memory coalescing behavior or shared memory bank conflicts. Ahn et al. [14] propose Trident, a GPU cache-based timing channel to recover all AES keys. Wang and Zhang [55] propose a profiling-based side-channel attack to fully recover the AES encryption secret key. Naghibijouybari et al. [41] develop several types of covert channels on different resources within a GPU. Nayak et al. [44] develop a similar microarchitectural covert channel on another resource, the GPU's shared last level translation lookaside buffer(TLB). Ahn et al. [15] exploit the contention on the GPU's on-chip interconnects (shared between SMs) to build microacrhitectural covert channels within discrete GPUs. Naghibijouybari et al. demonstrate a series of end-to-end GPU side channel attacks covering the different threat scenarios on both graphics and computational stacks, as well as across them [42, 43] in native applications. Dutta et al. [22] develop microarchitectural covert channels in Intel-based integrated CPU-GPU systems across the CPU and iGPU through shared LLC and on-chip ring interconnect. All of these microarchitectural attacks have been proposed in native applications.

## 9.3 Website Fingerprinting

Some of the discussed prior work in subsections 9.1 and 9.2 exploit microarchitectural side channel attacks to implement website fingerprinting. In Section 3, we conducted a thorough comparison of our threat model with these website fingerprinting attacks, which is presented in Table 1. We have also detailed the distinct characteristics of our attacks in comparison to these previous works in subsections 9.1 and 9.2.

For a comprehensive evaluation, we compare the classification accuracy between our attacks and those from other studies in Table 6. Our attack achieved precision in the closed-world setup that is comparable to the precision reported in [20]. However, in an open-world setup, our WebGPU-based attack outperformed theirs, reaching a success rate of 89.8% compared to their 81.4%. It is important to note that the target of the attack in [20] was ARM-based SoCs, which constitute a different platform than our work and other existing research. The comparison in Table 1 reveals that our WebGPU-Spy attack exhibits comparable performance to the native GPU attacks. Through the exploitation of WebGPU, which grants low-level access to the GPU's hardware, and by targeting the GPU's internal L3 cache, the attacker gains the ability to monitor the fine-grained leakage occurring within the GPU (where the rendering task is processed) through the compute stack.

## 10 CONCLUDING REMARKS

In this paper, we identified the GPU's internal caches (specifically L3 cache) as a new attack vector for remote microarchitectural attacks in web browsers. We exploited the separate memory access pathways of graphics and compute stacks to monitor the screen rendering process through the L3 cache using a compute spy and developed a basic cache occupancy channel within the GPU. Then, we optimized our attack by leveraging the GPU's inherent parallelism at different levels to achieve high-quality and high-resolution cache occupancy channel.

As computing systems are increasingly heterogeneous and acceleration APIs are becoming available through web interfaces, studying this new class of remote attacks is essential and critical to understanding how microarchitectural attacks manifest beyond just the CPU in widely used systems. We believe this work paves the way for future work studying the vulnerabilities exposed by the emerging web standards and helps us securely design these interfaces to reduce the threat posed by these attacks.

## REFERENCES
[1] 2015. Chromium: window.performance.now does not support sub-millisecond precision on Windows. https://bugs.chromium.org/p/chromium/issues/detail?id=158234#c110.
[2] 2020. Global PC GPU shipment share by vendor. https://www.statista.com/statistics/754557/worldwide-gpu-shipments-market-share-by-vendor/.
[3] 2022. CUDA, Nvidia. https://developer.nvidia.com/cuda-zone/.
[4] 2022. Mitigations landing for new class of timing attack. https://blog.mozilla.org/security/2018/01/03/mitigations-landing-new-class-timing-attack/.
[5] 2022. OpenCL Overview, Khronos Group. https://www.khronos.org/opencl/.

[6] 2022. OpenGL ES Overview, Khronos Group. https://www.khronos.org/opengles/.
[7] 2022. OpenGL Overview, Khronos Group. https://www.khronos.org/opengl/.
[8] 2022. performance.now. https://developer.mozilla.org/en-US/docs/Web/API/Performance/now.
[9] 2022. Site Isolation. https://www.chromium.org/Home/chromium-security/site-isolation/.
[10] 2022. Vulkan Overview, Khronos Group. https://www.khronos.org/vulkan/.
[11] 2022. WebGL Overview, Khronos Group. https://www.khronos.org/webgl/.
[12] 2022. WebGPU. https://www.w3.org/TR/webgpu/.
[13] 2023. WebGPU: Unlocking modern GPU access in the browser.
[14] Jaeguk Ahn, Cheolgyu Jin, Jiho Kim, Minsoo Rhu, Yunsi Fei, David Kaeli, and John Kim. 2021. Trident: A Hybrid Correlation-Collision GPU Cache Timing Attack for AES Key Recovery. In *2021 IEEE International Symposium on High-Performance Computer Architecture (HPCA)*. 332–344. https://doi.org/10.1109/HPCA51647.2021.00036
[15] Jaeguk Ahn, Jiho Kim, Hans Kasan, Leila Delshadtehrani, Wonjun Song, Ajay Joshi, and John Kim. 2021. Network-on-Chip Microarchitecture-Based Covert Channel in GPUs *(MICRO '21)*. Association for Computing Machinery, New York, NY, USA, 565–577. https://doi.org/10.1145/3466752.3480093
[16] Andrew Bortz and Dan Boneh. 2007. Exposing Private Information by Timing Web Applications. In *Proceedings of the 16th International Conference on World Wide Web* (Banff, Alberta, Canada) *(WWW '07)*. ACM, New York, NY, USA, 621–628. https://doi.org/10.1145/1242572.1242656
[17] Jie Chen and Guru Venkataramani. 2014. CC-Hunter: Uncovering Covert Timing Channels on Shared Processor Hardware. In *Proceedings of the International Symposium on Microarchitecture (MICRO)*.
[18] Alex Christensen. 2015. Reduce resolution of performance.now. https://bugs.webkit.org/show_bug.cgi?id=146531.
[19] David Cock, Qian Ge, Toby Murray, and Gernot Heiser. 2014. The Last Mile: An Empirical Study of Timing Channels on SeL4. In *Proceedings of the 2014 ACM SIGSAC Conference on Computer and Communications Security* (Scottsdale, Arizona, USA) *(CCS '14)*. Association for Computing Machinery, New York, NY, USA, 570–581. https://doi.org/10.1145/2660267.2660294
[20] Patrick Cronin, Xing Gao, Haining Wang, and Chase Cotton. 2021. An Exploration of ARM System-Level Cache and GPU Side Channels. In *Annual Computer Security Applications Conference* (Virtual Event, USA) *(ACSAC '21)*. Association for Computing Machinery, New York, NY, USA, 784–795. https://doi.org/10.1145/3485832.3485902
[21] Leonid Domnitser, Aamer Jaleel, Jason Loew, Nael Abu-Ghazaleh, and Dmitry Ponomarev. 2012. Non-monopolizable caches: Low-complexity mitigation of cache side channel attacks. *ACM Transactions on Architecture and Code Optimization* 8, 4 (2012). https://doi.org/10.1145/2086696.2086714
[22] Sankha Dutta, Hoda Naghibijouybari, Nael Abu-Ghazaleh, Andres Marquez, and Kevin Barker. 2021. Leaky Buddies: Cross-Component Covert Channels on Integrated CPU-GPU Systems. In *Proceedings of the International Symposium on Computer Architecture (ISCA)*.
[23] Dmitry Evtyushkin and Dmitry Ponomarev. 2016. Covert Channels through Random Number Generator: Mechanisms, Capacity Estimation and Mitigations. In *CCS*.
[24] Dmitry Evtyushkin, Dmitry Ponomarev, and Nael Abu-Ghazaleh. 2016. Understanding and mitigating covert channels through branch predictors. *ACM Transactions on Architecture and Code Optimization* 13, 1 (2016), 10.
[25] Pietro Frigo, Cristiano Giuffrida, Herbert Bos, and Kaveh Razavi. 2018. Grand Pwning Unit: Accelerating Microarchitectural Attacks with the GPU. In *Proceedings of IEEE Symposium on Security and Privacy*. 357–372. https://doi.org/10.1109/SP.2018.00022
[26] Ben Gras, Kaveh Razavi, Erik Bosman, Herbert Bos, and Cristiano Giuffrida. 2017. ASLR on the Line: Practical Cache Attacks on the MMU. In *NDSS*.
[27] Daniel Gruss, David Bidner, and Stefan Mangard. 2015. Practical Memory Deduplication Attacks in Sandboxed Javascript. In *Computer Security – ESORICS 2015*, Günther Pernul, Peter Y A Ryan, and Edgar Weippl (Eds.). Springer International Publishing, Cham, 108–122.
[28] Daniel Gruss, Raphael Spreitzer, and Stefan Mangard. 2015. Cache Template Attacks: Automating Attacks on Inclusive Last-Level Caches. In *24th USENIX Security Symposium (USENIX Security 15)*. USENIX Association, Washington, D.C., 897–912. https://www.usenix.org/conference/usenixsecurity15/technical-sessions/presentation/gruss
[29] Intel. 2015. *Intel Processor Graphics Gen9 Architecture*. Retrieved 2020 from https://software.intel.com/sites/default/files/managed/c5/9a/The-Compute-Architecture-of-Intel-Processor-Graphics-Gen9-v1d0.pdf
[30] Intel. 2019. *Intel Processor Graphics Gen11 Architecture*. Retrieved 2022 from https://software.intel.com/sites/default/files/managed/db/88/The-Architecture-of-Intel-Processor-Graphics-Gen11_R1new.pdf
[31] Aamer Jaleel, Eric Borch, Malini Bhandaru, Simon C. Steely Jr., and Joel Emer. 2010. Achieving Non-Inclusive Cache Performance with Inclusive Caches - Temporal Locality Aware (TLA) Cache Management Policies. In *Proceedings of the International Symposium on Microarchitecture (MICRO)*.
[32] Zhen Hang Jiang, Yunsi Fei, and David Kaeli. 2016. A complete key recovery timing attack on a GPU. In *2016 IEEE International Symposium on High Performance Computer Architecture (HPCA)*. 394–405. https://doi.org/10.1109/HPCA.2016.7446081
[33] Zhen Hang Jiang, Yunsi Fei, and David R. Kaeli. 2017. A Novel Side-Channel Timing Attack on GPUs. *Proceedings of the on Great Lakes Symposium on VLSI 2017* (2017).
[34] S. K. Khatamifard, L. Wang, S. Köse, and U. R. Karpuzcu. 2018. A New Class of Covert Channels Exploiting Power Management Vulnerabilities. *IEEE Computer Architecture Letters* 17, 2 (2018), 201–204.
[35] David Kohlbrenner and Hovav Shacham. 2016. Trusted Browsers for Uncertain Times. In *25th USENIX Security Symposium (USENIX Security 16)*. USENIX Association, Austin, TX, 463–480. https://www.usenix.org/conference/usenixsecurity16/technical-sessions/presentation/kohlbrenner
[36] Moritz Lipp, Daniel Gruss, Michael Schwarz, David Bidner, Clémentine Maurice, and Stefan Mangard. 2017. Practical Keystroke Timing Attacks in Sandboxed JavaScript. In *Computer Security – ESORICS 2017*, Simon N. Foley, Dieter Gollmann, and Einar Snekkenes (Eds.). Springer International Publishing, Cham, 191–209. https://link.springer.com/chapter/10.1007/978-3-319-66399-9_11
[37] Moritz Lipp, Daniel Gruss, Raphael Spreitzer, Clémentine Maurice, and Stefan Mangard. 2016. ARMageddon: Cache Attacks on Mobile Devices. In *25th USENIX Security Symposium (USENIX Security 16)*. USENIX Association, Austin, TX, 549–564. https://www.usenix.org/conference/usenixsecurity16/technical-sessions/presentation/lipp
[38] Fangfei Liu, Yuval Yarom, Qian Ge, Gernot Heiser, and Ruby B. Lee. 2015. Last-level cache side-channel attacks are practical. In *Security and Privacy (SP)*.
[39] Mozilla. 2018. SharedArrayBuffer. https://developer.mozilla.org/en-US/docs/Web/JavaScript/Reference/Global_Objects/SharedArrayBuffer.
[40] Mozilla. 2022. WebGL timer extension. https://developer.mozilla.org/en-US/docs/Web/API/EXT_disjoint_timer_query.
[41] Hoda Naghibijouybari, Khaled Khasawneh, and Nael Abu-Ghazaleh. 2017. Constructing and Characterizing Covert Channels on GPGPUs. In *Proceedings of the International Symposium on Microarchitecture (MICRO)*.
[42] Hoda Naghibijouybari, Ajaya Neupane, Zhiyun Qian, and Nael Abu-Ghazaleh. 2018. Rendered Insecure: GPU Side Channel Attacks Are Practical. In *Proceedings of the 2018 ACM SIGSAC Conference on Computer and Communications Security* (Toronto, Canada) *(CCS '18)*. ACM, New York, NY, USA, 2139–2153. https://doi.org/10.1145/3243734.3243831
[43] Hoda Naghibijouybari, Ajaya Neupane, Zhiyun Qian, and Nael Abu-Ghazaleh. 2021. Side Channel Attacks on GPUs. *IEEE Transactions on Dependable and Secure Computing* 18, 4 (2021), 1950–1961. https://doi.org/10.1109/TDSC.2019.2944624
[44] Ajay Nayak, Pratheek B., Vinod Ganapathy, and Arkaprava Basu. 2021. (Mis)Managed: A Novel TLB-Based Covert Channel on GPUs. In *Proceedings of the 2021 ACM Asia Conference on Computer and Communications Security* (Virtual Event, Hong Kong) *(ASIA CCS '21)*. Association for Computing Machinery, New York, NY, USA, 872–885. https://doi.org/10.1145/3433210.3453077
[45] Yossef Oren, Vasileios P. Kemerlis, Simha Sethumadhavan, and Angelos D. Keromytis. 2015. The Spy in the Sandbox: Practical Cache Attacks in JavaScript and Their Implications. In *Proceedings of the 22Nd ACM SIGSAC Conference on Computer and Communications Security* (Denver, Colorado, USA) *(CCS '15)*. ACM, New York, NY, USA, 1406–1418. https://doi.org/10.1145/2810103.2813708
[46] D. Page. 2005. Partitioned Cache Architecture as a Side-Channel Defense Mechanism. In *Crypt. ePrint Arch.*
[47] Colin Percival. 2005. Cache missing for fun and profit. In *BSDCan*.
[48] Moinuddin K. Qureshi and Yale N. Patt. 2006. Utility-Based Partitioning: A Low-Overhead, High-Performance, Runtime Mechanism to Partition Shared Caches. In *Proceedings of the International Symposium on Microarchitecture (MICRO)*.
[49] Thomas Ristenpart, Eran Tromer, Hovav Shacham, and Stefan Savage. 2009. Hey, You, Get off of My Cloud: Exploring Information Leakage in Third-Party Compute Clouds. In *Proceedings of the 16th ACM Conference on Computer and Communications Security* (Chicago, Illinois, USA) *(CCS '09)*. Association for Computing Machinery, New York, NY, USA, 199–212. https://doi.org/10.1145/1653662.1653687
[50] Michael Schwarz, Clémentine Maurice, Daniel Gruss, and Stefan Mangard. 2017. Fantastic Timers and Where to Find Them: High-Resolution Microarchitectural Attacks in JavaScript. In *Financial Cryptography and Data Security*, Aggelos Kiayias (Ed.). Springer International Publishing, Cham, 247–267.
[51] Anatoly Shusterman, Ayush Agarwal, Sioli OConnell, Daniel Genkin, Yossi Oren, and Yuval Yarom. 2021. Prime+Probe 1, JavaScript 0: Overcoming Browser-based Side-Channel Defenses. In *USENIX Security Symposium, 2021*.
[52] Anatoly Shusterman, Lachlan Kang, Yarden Haskal, Yosef Meltser, Prateek Mittal, Yossi Oren, and Yuval Yarom. 2019. Robust Website Fingerprinting Through the Cache Occupancy Channel. In *28th {USENIX} Security Symposium ({USENIX} Security 19)*. 639–656.
[53] Antonin Durey Vitaly Dyadyuk Pierre Laperdrix et al. Tomer Laor, Naif Mehanna. 2022. DRAWNAPART: A Device Identification Technique based on Remote GPU Fingerprinting. In *Network and Distributed System Security Symposium* (San Diego).

[54] Tom Van Goethem, Wouter Joosen, and Nick Nikiforakis. 2015. The Clock is Still Ticking: Timing Attacks in the Modern Web. In *Proceedings of the 22Nd ACM SIGSAC Conference on Computer and Communications Security* (Denver, Colorado, USA) *(CCS '15)*. ACM, New York, NY, USA, 1382–1393. https://doi.org/10.1145/2810103.2813632

[55] Xin Wang and Wei Zhang. 2020. An Efficient Profiling-Based Side-Channel Attack on Graphics Processing Units. In *National Cyber Summit (NCS) Research Track*, Kim-Kwang Raymond Choo, Thomas H. Morris, and Gilbert L. Peterson (Eds.). Springer International Publishing, Cham, 126–139.

[56] Zhenghong Wang and Ruby B Lee. 2006. Covert and Side Channels Due to Processor Architecture.. In *Computer Security Applications Conference (ACSAC)*.

[57] Shujiang Wu, Jianjia Yu, Min Yang, and Yinzhi Cao. 2022. Rendering Contention Channel Made Practical in Web Browsers. In *31st USENIX Security Symposium (USENIX Security 22)*. USENIX Association, Boston, MA, 3183–3199. https://www.usenix.org/conference/usenixsecurity22/presentation/wu-shujiang

[58] Qiumin Xu, Hoda Naghibijouybari, Shibo Wang, Nael Abu-Ghazaleh, and Murali Annavaram. 2019. GPUGuard: Mitigating Contention Based Side and Covert Channel Attacks on GPUs. In *Proceedings of the ACM International Conference on Supercomputing* (Phoenix, Arizona) *(ICS '19)*. ACM, New York, NY, USA, 497–509. https://doi.org/10.1145/3330345.3330389

[59] Fan Yao, Milos Doroslovacki, and Guru Venkataramani. 2018. Are Coherence Protocol States Vulnerable to Information Leakage?. In *Proceedings of the International Symposium on High Performance Computer Architecture (HPCA)*.

[60] Yuval Yarom and Katrina Falkner. 2014. FLUSH+RELOAD: A High Resolution, Low Noise, L3 Cache Side-Channel Attack. In *23rd USENIX Security Symposium (USENIX Security 14)*. USENIX Association, San Diego, CA, 719–732. https://www.usenix.org/conference/usenixsecurity14/technical-sessions/presentation/yarom

[61] Michael K. Reiter Yinqian Zhang, Ari Juels and Thomas Ristenpart. 2014. Cross-Tenant Side-Channel Attacks in PaaS Clouds. In *Proceedings of the 2014 ACM SIGSAC Conference on Computer and Communications Security*. 990–1003. https://doi.org/10.1145/2660267.2660356

[62] Boris Zbarsky. 2015. Reduce resolution of performance.now. https://hg.mozilla.org/integration/mozilla-inbound/rev/48ae8b5e62ab.